\documentclass[a4paper,11pt]{article}
\usepackage{pos}

\usepackage{graphicx}
\usepackage{amsmath}
\usepackage[dvipsnames]{xcolor}
\usepackage{hyperref}
\usepackage{cleveref}
\usepackage{booktabs}
\usepackage{physics}
\usepackage{dsfont}
\usepackage{tcolorbox}
\usepackage{enumitem}
\usepackage{graphicx}
\usepackage{wrapfig}
\usepackage[export]{adjustbox}
\usepackage{setspace}
\usepackage{bbm}
\usepackage{bm}
\usepackage{orcidlink}
\usepackage{soul}
\usepackage{subcaption}
\usepackage{array}
\usepackage{multirow}
\usepackage[table]{xcolor}
\usepackage{caption}
\usepackage{soul}

\usepackage{tikz}
\usepackage[compat = 1.1.0]{tikz-feynman}
\usepackage[customcolors]{hf-tikz}
\usetikzlibrary{positioning,calc,intersections,patterns.meta}

\usepackage{xcolor}
\definecolor{Green}{RGB}{0,150,0}

\newlength{\QCDvertA}
\newlength{\QCDvertD}

\NewDocumentCommand\semiloop{O{black}mmmO{}O{above}}
{%
\draw[#1] let \p1 = ($(#3)-(#2)$) in (#3) arc (#4:({#4+180}):({0.5*veclen(\x1,\y1)})node[midway, #6] {#5};)
}

\newcommand{\Seun}[3]{
\begin{tikzpicture}[dot/.style={draw,circle,minimum size=2.5pt,inner sep=0pt,outer sep=0pt,fill=Green, color=Green}]
\begin{feynman}
\vertex (n1) ;
\vertex[right=\QCDvertA of n1] (n2) ;

\vertex at ($(n1)!0.17!(n2)$) (m1);
\vertex at ($(n2)!0.17!(n1)$) (m2);

\vertex[above=2.3*\QCDvertD of m1] (m3);
\vertex[above=2.3*\QCDvertD of m2] (m4);

\path[name path=Aline] (n1) to[in=+120, out=+60] (n2);
\path[name path=Bline] (m1) -- (m3);
\path[name path=Cline] (m2) -- (m4);

\path[name intersections={of=Aline and Bline}];
\vertex at (intersection-1) (c1);

\path[name intersections={of=Aline and Cline}];
\vertex at (intersection-1) (c2);

\diagram*{
        (n1) -- [fermion, in = +120, out = +60, color=#1 ] (n2) ,
        (n1) -- [fermion,                     , color=#2 ] (n2) ,
        (n1) -- [fermion, in = -120, out = -60, color=#3  ] (n2) ,
        (c1) -- [photon,  in = +100, out = +80, color=Green] (c2) ,
    };

\node [dot] at (c1);
\node [dot] at (c2);
\end{feynman}
\end{tikzpicture}
}

\newcommand{\Exun}[3]{
\begin{tikzpicture}[dot/.style={draw,circle,minimum size=2.5pt,inner sep=0pt,outer sep=0pt,fill=Green, color=Green}]
\begin{feynman}
\vertex (n1) ;
\vertex[right=\QCDvertA of n1] (n2) ;

\vertex at ($(n1)!0.17!(n2)$) (m1);
\vertex at ($(n2)!0.17!(n1)$) (c2);

\vertex[above=2.3*\QCDvertD of m1] (m3);

\path[name path=Aline] (n1) to[in=+120, out=+60] (n2);
\path[name path=Bline] (m1) -- (m3);

\path[name intersections={of=Aline and Bline}];
\vertex at (intersection-1) (c1);

\diagram*{
        (n1) -- [fermion, in = +120, out = +60, color=#1 ] (n2) ,
        (n1) -- [fermion,                     , color=#2 ] (n2) ,
        (n1) -- [fermion, in = -120, out = -60, color=#3 ] (n2) ,
        (c1) -- [photon,                      , color=Green] (c2) ,

    };
\node [dot] at (c1);
\node [dot] at (c2);
\end{feynman}
\end{tikzpicture}
}

\newcommand{\mcrun}[3]{
\begin{tikzpicture}[
dot/.style={draw,circle,minimum size=2.5pt,inner sep=0pt,outer sep=0pt,fill=Green, color=Green},
square/.style={rectangle,minimum size=6pt,draw=Green, outer sep=6pt, fill=Green, pattern={Hatch[angle=45,distance=1.2pt,xshift=.1pt]}, pattern color=Green!70!black}]

\begin{feynman}
\vertex (n1) ;
\vertex[right=\QCDvertA of n1] (n2) ;

\vertex at ($(n1)!0.50!(n2)$) (i1);
\vertex[above=2.3*\QCDvertD of i1]   (i2);

\path[name path=Aline] (n1) to[in=+120, out=+60] (n2);
\path[name path=Bline] (i1) -- (i2);
\path[name intersections={of=Aline and Bline}];
\vertex at (intersection-1) (c1);

\vertex at ($(n1)!0.54!(n2)$) (i1);
\vertex[above=0.6cm of i1]   (i2);

    \diagram*{
        (n1) -- [fermion, in = +180, out = +60, color=#1 ] (c1) ,
        (c1) -- [fermion, in = +120, out =   0, color=#1 ] (n2) ,
        (n1) -- [fermion,                     , color=#2 ] (n2) ,
        (n1) -- [fermion, in = -120, out = -60, color=#3 ] (n2) ,
    };
\node[square] at (c1);
\end{feynman}
\end{tikzpicture}
}

\newcommand{\seabub}[3]{
\begin{tikzpicture}[
dot/.style={draw,circle,minimum size=2.5pt,inner sep=0pt,outer sep=0pt,fill=Green, color=Green},
square/.style={rectangle,minimum size=6pt,draw=Green, outer sep=6pt, fill=Green, pattern={Hatch[angle=45,distance=1.2pt,xshift=.1pt]}, pattern color=Green!70!black}]
]

\begin{feynman}
\vertex (n1) ;
\vertex[right=\QCDvertA of n1] (n2) ;

\vertex at ($(n2) + (0.5*\QCDvertD, 0.6*\QCDvertD)$) (c1);
\vertex at ($(n2) + (0.5*\QCDvertD, 1.4*\QCDvertD)$) (c2);

\vertex at ($(n1)!0.750!(n2) + (0, 0.6*\QCDvertD + 0.35cm)$) (c3);
\vertex at ($(n1)!0.750!(n2) + (0, 1.4*\QCDvertD + 0.35cm)$) (c4);
\vertex at ($(n1)!0.750!(n2) + (0.4*\QCDvertD, 1.0*\QCDvertD + 0.35cm)$) (c5);

\diagram*{
        (n1) -- [fermion, in = +120, out = +60, color=#1 ] (n2) ,
        (n1) -- [fermion,                     , color=#2 ] (n2) ,
        (n1) -- [fermion, in = -120, out = -60, color=#3  ] (n2) ,
        (c1) -- [photon, thick, color=Green ] (c2) ,
    };

\semiloop[]{c2}{c1}{-90}[][right];
\semiloop[fermion]{c1}{c2}{90};

\semiloop[]{c4}{c3}{-90}[][right];
\semiloop[fermion]{c3}{c4}{90};

\node[dot] at (c1);
\node[dot] at (c2);

\node[square] at (c5);

\node at ($(n1)!0.95!(n2) + (0, 1.4*\QCDvertD)$) {\(\Big/\)};

\end{feynman}
\end{tikzpicture}}

\newcommand{\seaval}[3]{
\begin{tikzpicture}[
dot/.style={draw,circle,minimum size=2.5pt,inner sep=0pt,outer sep=0pt,fill=Green, color=Green},
square/.style={rectangle,minimum size=6pt,draw=Green, outer sep=6pt, fill=Green, pattern={Hatch[angle=45,distance=1.2pt,xshift=.1pt]}, pattern color=Green!70!black}]
]

\begin{feynman}
\vertex (n1) ;
\vertex[right=\QCDvertA of n1] (n2) ;

\vertex at ($(n1)!0.17!(n2)$) (m1);
\vertex[above=2.3*\QCDvertD of m1] (m2);

\path[name path=Aline] (n1) to[in=+120, out=+60] (n2);
\path[name path=Bline] (m1) -- (m2);

\path[name intersections={of=Aline and Bline}];
\vertex at (intersection-1) (m1);

\vertex at ($(n2) + (-0.2*\QCDvertD, 0.8*\QCDvertD)$) (c1);
\vertex at ($(n2) + (-0.2*\QCDvertD, 1.6*\QCDvertD)$) (c2);

\diagram*{
        (n1) -- [fermion, in = +120, out = +60, color=#1 ] (n2) ,
        (n1) -- [fermion,                     , color=#2 ] (n2) ,
        (n1) -- [fermion, in = -120, out = -60, color=#3  ] (n2) ,
        (m1) -- [photon, thick, bend left, color=Green ] (c2) ,
    };

\semiloop[]{c2}{c1}{-90}[][right];
\semiloop[fermion]{c1}{c2}{90};

\node[dot] at (c2);
\node[dot] at (m1);

\end{feynman}
\end{tikzpicture}}
\usetikzlibrary{backgrounds}

\definecolor{llbl}{HTML}{DBE5F1}
\newcommand{\lcline}[1]{\arrayrulecolor{lightgray}\cline{#1}\arrayrulecolor{black}}
\newcommand{\timd}[1]{\mathcal{D}_m{#1}}

\let\OLDthebibliography\thebibliography
\renewcommand\thebibliography[1]{
  \OLDthebibliography{#1}
  \setlength{\parskip}{1pt}
  \setlength{\itemsep}{0.4ex plus 0.2ex}
}

\newcommand{\smallcap}[1]{{\mathchoice
  {\scalebox{1.1}{$\scriptstyle \mathsf{#1}$}}
  {\scalebox{1.1}{$\scriptstyle \mathsf{#1}$}}
  {\scalebox{1.1}{$\scriptscriptstyle \mathsf{#1}$}}
  {\scalebox{1.1}{$\scriptscriptstyle \mathsf{#1}$}}
}}

\newcommand{\EM}[0]{\smallcap{\rm EM}}
\newcommand{\iso}[0]{\smallcap{\rm isoQCD}}
\newcommand{\pp}[0]{\phantom{+}}

\title{Measuring Mass Splittings in Baryon SU(2) Multiplets}

\author[a,b]{Constantia Alexandrou}
\author[a]{Simone Bacchio}
\author[b]{Antonio Evangelista}
\author[a,b]{~~~~~~~~~~~Christos Iona}
\author*[b]{Christian Schneider}

\affiliation[a]{Computation-based Science and Technology Research Center, The Cyprus Institute, 20 Kavafi Str., Nicosia 2121, Cyprus}
\affiliation[b]{Department of Physics, University of Cyprus, P.O. Box 20537, 1678 Nicosia, Cyprus}

\emailAdd{schneider.christian@ucy.ac.cy}

\abstract{The inclusion of isospin-breaking corrections arising from electromagnetic interactions and the up–down quark mass difference is becoming increasingly important in  lattice QCD calculations, as the precision of observables is at or approaches the percent level. In this context, the RM123 method has been employed  in recent years to take into account such isospin corrections. In this work, we  study baryon mass splittings using the RM123 approach. Preliminary results are presented for the mass splittings within spin-1/2 and spin-3/2 baryon SU(2) multiplets, obtained using twisted-mass ensembles tuned close to physical quark masses. We consider only valence contributions, with the splittings computed at leading order in the isospin-breaking parameters.}

\FullConference{
The 43rd International Symposium on Lattice Field Theory (Lattice 2026)
}

\begin{document}

\maketitle

\setlength{\parindent}{0pt}
\setlength{\parskip}{6pt}

\section{Introduction and motivation}
Lattice QCD calculations are generally performed in isospin-symmetric QCD (isoQCD), where the up- and down-quark masses are taken to be degenerate and electromagnetic interactions are neglected. This setup avoids the additional complications associated with including QED on the lattice, such as the treatment of the photon zero mode and the long-range nature of the electromagnetic interaction. These effects are expected to be at the percent level or smaller for many observables, and are therefore negligible when the target precision does not reach that level.

A direct consequence of this setup is that hadrons belonging to the same isospin multiplet are exactly degenerate. In nature, however, these degeneracies are lifted by isospin-breaking effects, with mass splittings typically of $\mathcal{O}(5~\mathrm{MeV})$, corresponding to deviations of several percent for the pion mass. As the precision of current lattice calculations reaches the percent level and better, leading-order strong isospin-breaking (SIB) and electromagnetic (QED) corrections can no longer be neglected and must therefore be incorporated when going from isoQCD to QCD+QED. The corrections connecting these two theories are then computed as a perturbative expansion in the electromagnetic coupling, $\alpha_{\EM}$, and the up--down quark mass difference, $\delta_{ud}{=}(\mu_u{-}\mu_d)/\Lambda_{\smallcap{\rm QCD}}$. Restricting the calculation to the leading terms in this expansion constitutes the RM123 approach~\cite{deDivitiis:2011eh,deDivitiis:2013xla}, which provides a systematic framework for computing isospin-breaking effects without the need for fully dynamical QCD+QED simulations. 

Hadron mass splittings provide an especially clean setting in which to study these effects, particularly within the RM123 framework: they are experimentally well measured; and they isolate isospin-breaking contributions at the theoretical level, since the isospin-symmetric contributions cancel by construction. Moreover, the computationally challenging disconnected contributions that enter individual hadronic observables cancel in the corresponding mass splittings when the leading-order corrections are computed around the isoQCD point~\cite{Giusti:2017dmp}. These features make hadron mass splittings particularly well suited both for determining the relevant scale-setting parameters and for benchmarking the resulting leading-order isospin-breaking corrections.

\section{From isoQCD to QCD+QED} \label{sec:QCD+QED}

We follow the Edinburgh/FLAG prescription~\cite{FlavourLatticeAveragingGroupFLAG:2024oxs}, which defines an $N_{\!f}{=}2{+}1{+}1$ isoQCD theory through a meson-based scheme. Specifically, we fix the four parameters of the theory by requiring that, at every lattice spacing and, consequently, in the continuum limit~\cite{ExtendedTwistedMass:2024nyi},
\begin{equation}
\big\{g,\,\mu_\ell,\,\mu_s,\,\mu_c\big\}_{\iso}~~\Longrightarrow~~
\big\{f_{\pi},\,m_{\pi},\,m_{K},\,m_{D_s}\big\}\Big|_{\iso} \!= \big\{130.5,\,135,\,494.6,\,1967\big\}~\text{MeV}.
\end{equation}
Alternative scale-setting prescriptions, as well as different choices of the quantities used to define the isoQCD theory, produce observables that differ at the level of isospin-breaking effects.
In contrast, once QCD+QED is considered, these parameters, supplemented by $\delta_{ud}$ and $\alpha_{\EM}$, define, according to the Standard Model, a unique theory for the hadronic sector. Consequently, hadronic observables should no longer depend on the choice of scale-setting prescription. Within the RM123 framework, we work at leading order in $\alpha_{\EM}$, for which we set $\alpha_{\EM}{\approx}1/137$. At this order, the $U(1)$ gauge fields are independent of the SU(3) gauge fields and they are generated according to the $\mathrm{QED}_L$ prescription~\cite{Hayakawa:2008an,Davoudi:2018qpl} in Coulomb gauge, while the quarks couple to both gauge fields through the Dirac operator. We then determine the up-, down-, strange-, and charm-quark masses by requiring the masses of the $\pi^+$, $K^+$, $K^0$, and $D_s$ mesons to reproduce their experimental values~\cite{Giusti:2017dmp,Evangelista:2025dzi}.

Using the RM123 method, the expectation value of a hadron mass in QCD+QED can then be approximated at leading order as
\begin{equation}
m^{\smallcap{\mathrm{QCD+QED}}} \equiv  m^{\iso} + \Delta m|_{\iso} = m^{\iso} + \alpha_{\EM} \left.\pdv{m}{\alpha}\right|_{\iso} 
\!\!\!+
\Delta \vec{\mu} \left.\pdv{m}{\vec{\mu}}\right|_{\iso} \!\!\!+ \order{\alpha_{\EM}^2, \Delta\vec{\mu}^2, \alpha_{\EM} \Delta\vec{\mu}}
\label{eq:exp}
\end{equation}
where $\Delta\vec{\mu}{=}\{\Delta\mu_u,\Delta\mu_d,\Delta\mu_s,\Delta\mu_c\}$ denotes the  shifts in the quark masses from isoQCD to QCD+QED, according to the prescriptions described above. The mass correction $\Delta m|_{\iso}$ can be extracted by studying the large Euclidean-time behaviour of suitably constructed ratios of correlation functions as follows. Let $C(t)$, with parameters $\vec{x}_0$, denote the correlator at the isoQCD point, and $C'_x(t)$ denote the correlator obtained by infinitesimally varying the corresponding parameter, $x_0 \to x_0 + \delta x$. Considering for simplicity only the ground-state contribution, we write
\begin{equation}
C(t)\approx Ae^{-mt}
\qq{and}
C'_x(t) =
C(t) + \delta_{\!x} C(t)
\approx
(A{+}\delta_{\!x} A)e^{-(m+\delta_{\!x} m)t}\,,
\end{equation}
and the mass derivative is then extracted by performing the following procedure:
\begin{equation}
X(t,\delta x) \equiv \frac{1}{\delta x} \frac{\delta_{\!x} C(t)}{C(t)} \approx \frac{1}{\delta x} \left(\frac{\delta_{\!x} A}{A}-\delta_{\!x} m\,t\right)\, \quad \Longrightarrow \quad
\partial_x m= \mathcal{D}_m X \equiv - \lim_{\substack{t \to \infty \\ \delta x \to 0}}\partial_t X(t,\delta x)\,.
\label{eq:LOexp}
\end{equation}
We remark that in  Eq.~\eqref{eq:exp}, we require a derivative with respect to (w.r.t.) $\alpha$, which reads as a second derivative w.r.t. the electric charge, and thus we perform the following procedure in that case:
\begin{equation}
    \alpha\frac{\partial}{\partial\alpha} = \frac{e^2}{2}\frac{\partial^2}{\partial e^2} \quad\Longrightarrow\quad \alpha\frac{\partial m}{\partial \alpha} = e^2 \mathcal{D}_m E\qq{with}E(t,\delta e)=\frac{C_{+e}(t)+C_{-e}(t)-2C(t)}{2\,(\delta e)^2\,C(t)}\,,
\end{equation}
where the latter isolates via finite difference the second derivative in $e$.

An additional consideration arising from the twisted mass regularisation employed in this work is that the inclusion of QED leads to two distinct critical hopping parameters for the maximal twist condition: $\kappa^{\mathrm{cr}}_{u}$ for the positively charged quarks ($u,c$), with charge $e_u{=}+2/3 e$, and $\kappa^{\mathrm{cr}}_{d}$ for the negatively charged quarks ($d,s$), with charge $e_d{=}-1/3 e$. Since tuning to maximal twist is necessary to ensure automatic $O(a)$ improvement~\cite{Frezzotti:2000nk, Frezzotti:2003ni}, we determine both critical hopping parameters by requiring simultaneously the restoration of the PCAC relation and parity~\cite{Frezzotti:2016lwv}. 

\section{SU(2) mass splittings}

\begin{figure}[tbh]
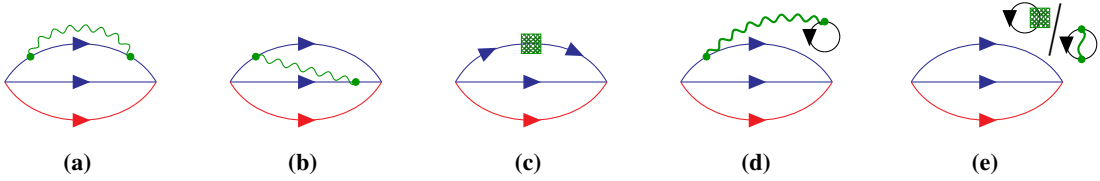

\centering
\vspace{-1em}
\pgfmathsetlength{\QCDvertA}{2.0cm}
\pgfmathsetlength{\QCDvertD}{0.5cm}
\subfloat[\label{sub:a}]{$\vcenter{\hbox{\Seun{Blue}{Blue}{Red}}}$}
\qquad
\subfloat[\label{sub:b}]{$\vcenter{\hbox{\Exun{Blue}{Blue}{Red}}}$}
\qquad
\subfloat[\label{sub:c}]{$\vcenter{\hbox{\mcrun{Blue}{Blue}{Red}}}$}
\qquad
\subfloat[\label{sub:d}]{$\vcenter{\hbox{\seaval{Blue}{Blue}{Red}}}\!\!\!$}
\qquad
\subfloat[\label{sub:e}]{$\vcenter{\hbox{\seabub{Blue}{Blue}{Red}}}\!\!\!\!\!\!\!\!\!$}
\caption{
Examples of the diagrams contributing at leading order to the baryon two-point correlation function through the up--down quark-mass splitting and the electromagnetic coupling. These include the valence corrections—self-energy (a), photon exchange (b), and quark-mass corrections (c)—as well as electromagnetic sea--valence (d) and purely sea (e) contributions.
}
\label{fig:baryon_diagrams}
\end{figure}

The key advantage of considering mass splittings associated with SU(2) symmetry breaking within hadron multiplets is that, at leading order, they isolate pure isospin-breaking effects. These receive contributions only from purely valence terms, depicted by diagrams (a)--(c) in Fig.~\ref{fig:baryon_diagrams}, and from electromagnetic sea--valence interactions, diagram (d), in which a sea-quark loop interacts electromagnetically with a valence quark. At  leading-order in the expansion around the isoQCD point, the isoQCD mass $m_B^{\iso}$ cancels in a mass splitting  such as that of the nucleon, $\Delta m_N {\equiv}  m_n {-} m_p$, since it contributes equally to both the neutron $m_n$ and proton $m_p$ masses in Eq.~\eqref{eq:exp}. Moreover, the purely disconnected contributions, depicted in diagram~(e), also cancel exactly, since they enter symmetrically at leading order and are evaluated in the isoQCD limit. Similarly, for baryons containing strange or charm quarks, these additional flavours act as spectators, entering symmetrically in the two members of the multiplet. Their mass difference in isoQCD and QCD+QED therefore does not contribute to the splitting.

We demonstrate our procedure by detailing the description for the case of the nucleon splitting. We restrict our analysis to the purely valence contributions, which are expected to be the dominant ones. The corrections to the neutron and proton mass, respectively, are
\begin{align}
\Delta m_{n} & =  \timd{\Big\{
e_u^{2}E_{uu}^n + e_u e_d E_{ud}^n\, + e_d^2 E_{dd}^n\,
+ 
\Delta \kappa^{\mathrm{cr}}_u P_{u}^n + \Delta \kappa^{\mathrm{cr}}_d P_{d}^n
+
\Delta \mu_u S_{u}^n + \Delta \mu_d S_{d}^n
\Big\}}\,,
\\
\Delta m_{p} & = \timd{\Big\{
e_u^2 E_{uu}^p\, + e_u e_d E_{ud}^p\, + e_d^{2}E_{dd}^p
+ 
\Delta \kappa^{\mathrm{cr}}_{u} P_{u}^p + \Delta \kappa^{\mathrm{cr}}_d P_{d}^p
+
\Delta \mu_u S_{u}^p + \Delta \mu_d S_{d}^p
\Big\}}\,.
\end{align}
Here, the terms $E_{f_1f_2}^{B}$ denote the second derivatives with respect to the electric charge, involving the flavours $f_1$ and $f_2$. The terms $P_f^{B}$ denote derivatives with respect to the hopping parameter $\kappa$, yielding a pseudoscalar current acting on the flavour $f$, while $S_f^{B}$ denote derivatives with respect to the quark mass yielding a scalar current. Furthermore, since these quantities are evaluated in isoQCD, we can exploit the $u{\to}d$ and $n{\to}p$ symmetries. For example, $E_{uu}^{n}=E_{dd}^{p}\equiv E_{dd}^{N}$ and $P/S_{u}^{n}=P/S_{d}^{p}\equiv P/S_{d}^{N}$.
Using these relations, one obtains
\begin{align}
\Delta m_N & = \timd{\bigg\{
\underbrace{-\frac{e^2}{3}\,\left[E_{uu}^{N} - E_{dd}^{N} \right]
+\Delta \kappa^{\mathrm{cr}}_{ud}
\left[P_{u}^{N} - P_{d}^{N} \right]}_{\rm QED}
\,+\,\underbrace{\phantom{\frac11}\!\!\!\!\!\Delta \mu_{ud} \left[S_{u}^{N} - S_{d}^{N} \right]}_{\rm SIB}
\bigg\}}
\label{eq:nucleon_MS}
\end{align}
where $\kappa^{\mathrm{cr}}_{ud} = \kappa^{\mathrm{cr}}_{d} - \kappa^{\mathrm{cr}}_{u}$ and $\Delta \mu_{ud} = \Delta \mu_{d} - \Delta \mu_{u}$. Since the splitting $\kappa^{\mathrm{cr}}_{ud}$ arises as a QED-induced counterterm and is strongly statistically correlated with the electric-charge derivatives of the correlators, we combine these contributions in the following analysis and refer to their sum as the QED contribution to the mass correction. In the final results, we additionally include the universal QED-induced finite-volume effects (FVE), which constitute one of the dominant systematic uncertainties in calculations of radiative corrections. These effects arise from the absence of a mass gap in the QCD+QED spectrum and consequently scale polynomially with inverse powers of the lattice extent $L$. As shown in Ref.~\cite{BMW:2014pzb}, the leading and next-to-leading corrections in $1/L$ are universal. Specifically, the finite-volume hadron mass $m_B(L)$ is related to its infinite-volume value $m_B$ by
\begin{equation}
m_B(L) = m_B \left\{ 1 - e_B^2\;\frac{c_{\smallcap{EM}}}{8\pi} \left( \frac{1}{m_BL} + \frac{2}{m_B^2 L^2} \right) + \order{L^{-3}} \right\} 
\qq{with} c_{\smallcap{EM}} = 2.837297\,(1) \, ,
\label{eq:FVcorrection}
\end{equation}
where $e_B$ is the charge of baryon and $c_{\smallcap{EM}}$ is the result of an analytic integral~\cite{BMW:2014pzb}.

\section{Preliminary Results}

In this work, we present a first analysis based on two ETMC ensembles tuned at quark masses close to those matching the Edinburg/FLAG scheme~\cite{ExtendedTwistedMass:2024nyi},  properties of which are summarised in Table~\ref{tab:ensembles}. Both ensembles have $N_f{=}2{+}1{+}1$ dynamical flavours and a similar spatial extent, with $L\approx5.5$~fm, but differ in the lattice spacing: ensemble B64 has $a{\approx}0.08$\,fm and ensemble D96  $a{\approx}0.06$\,fm. Two-point functions are computed using spin-colour-diluted point sources, while the photon fields are generated according to the QED$_{L}$ prescription in Coulomb gauge. The expectation value over the photon fields is then obtained by sampling them independently for each source and configuration.

\begin{table}[ht]
\centering
\renewcommand{\arraystretch}{1.2}
\aboverulesep=0ex % Solution part 1 of 3
\belowrulesep=0ex % Solution part 1 of 3
\begin{tabular}{cccc|ccc}
\toprule
Ensemble & $L\slash a$ & $a~[\text{fm}]$  & $L~[\text{fm}]$  & $a\mu^{\iso}$ & $a\Delta\mu_{ud}$ & $\Delta\kappa_{ud}^{\rm cr}$\\
\midrule
B64 & $64$ & 0.07948(11) & 5.09 & $0.000667(3)$ & $0.000512(3)$  & $-0.0003018(1)$    \\
D96 & $96$ & 0.05685(09) & 5.46 & $0.000493(2)$ & $0.000371(1)$  & $-0.0002928(1)$   \\
\bottomrule
\end{tabular}
\caption{Parameters of the ETMC ensembles used in this work with scale setting described in Refs.~\cite{ExtendedTwistedMass:2024nyi, Evangelista:2025dzi}.}
\label{tab:ensembles}
\end{table}

As discussed in the previous section, to extract the mass splittings, we first combine all relevant two-point functions to construct the QED and SIB contributions, see, e.g., Eq.~\eqref{eq:nucleon_MS} for $\Delta m_N$. We then study the large-distance behaviour of suitably constructed ratios, as in Eq.~\eqref{eq:LOexp}, from which the mass splittings are extracted. This procedure preserves the correlations among the different contributions, thereby avoiding a loss of statistical information. To improve the signal-to-noise ratio, which deteriorates significantly towards the physical point, we use several valence light-quark masses heavier than the isoQCD value. Specifically, we consider quark masses ranging from the physical light-quark mass up to the strange-quark mass, with the latter included as the upper endpoint. This range also allows us to investigate SU(3) isospin-breaking effects. We also increase the number of sources as we approach the light quark mass to keep statistical errors roughly constant. Finally, we extrapolate the results in the valence quark masses; the current data indicate that a linear dependence provides an adequate description.

In Fig.~\ref{fig:baryon_diagrams}, we show our analysis for the $\Sigma$ baryon using the B64 ensemble.  The correlator exhibits a clear dependence on the valence-quark mass. The top panels show the QED and SIB contributions, while the lower panels display the forward lattice-time derivative of the corresponding ratios. At each valence mass, we extract the large time limit of Eq.~\eqref{eq:LOexp} by identifying a plateau region in which excited states are  sufficiently suppressed. A comprehensive assessment of  associated systematic uncertainties, including the dependence on fit intervals and   the degree of excited state suppression, will be quantified in future analyses.
\begin{center}
\includegraphics[width=\linewidth]{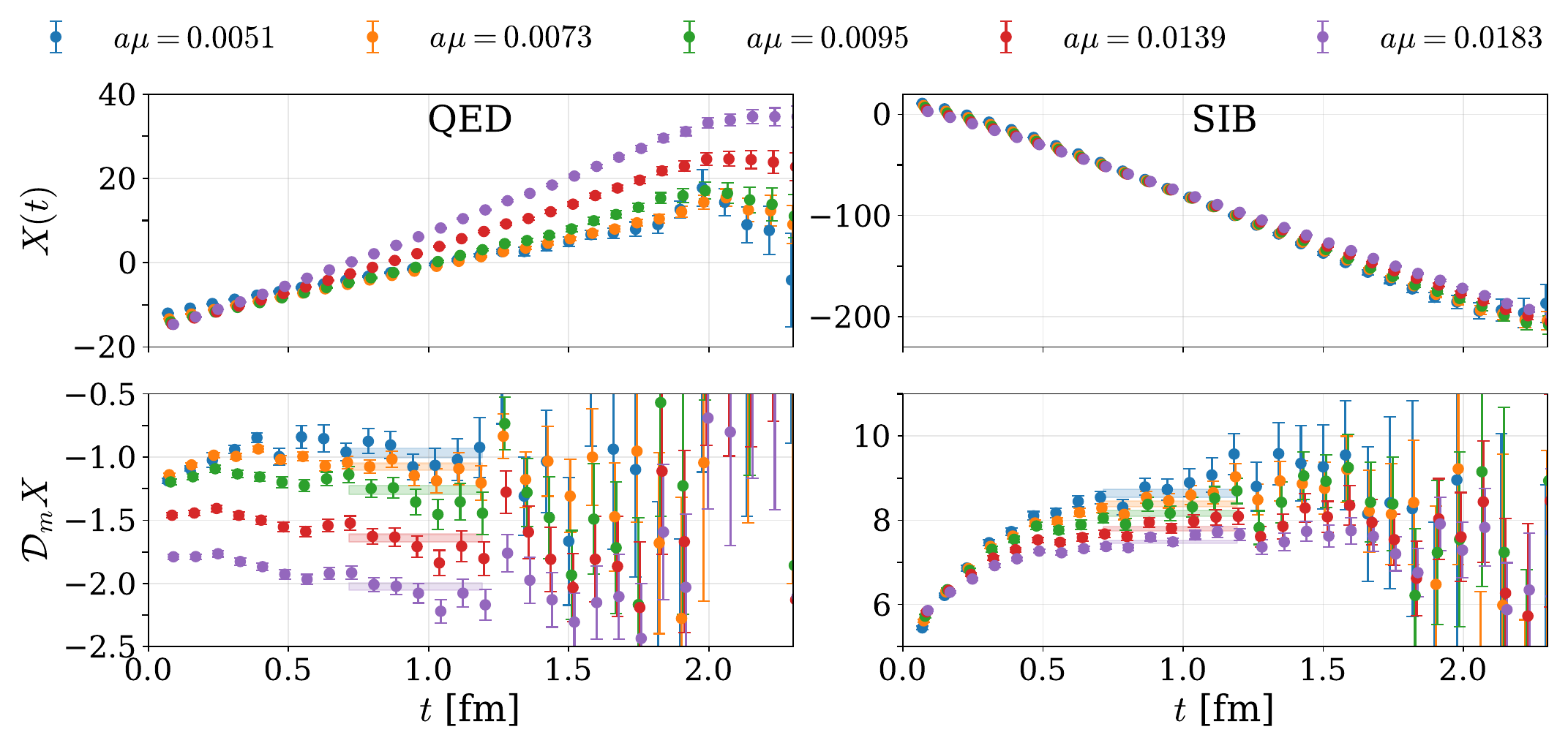}
\captionof{figure}{In the top panels, we show   the time dependence of the correlation functions computed using the B64 ensemble for the QED (left) and SIB (right) ratios in Eq.~\eqref{eq:LOexp} for the $\Sigma$ baryon, $m_{\Sigma^-}-m_{\Sigma^+}$. The bottom panels show the forward-time derivatives of the these ratios, with the bands representing the results of plateau fits. Each colour corresponds to a different valence quark mass, as indicated in the header of the figure.\label{fig:SlopeExtraction}}
\vspace{2em}
\includegraphics[width=0.9\linewidth]{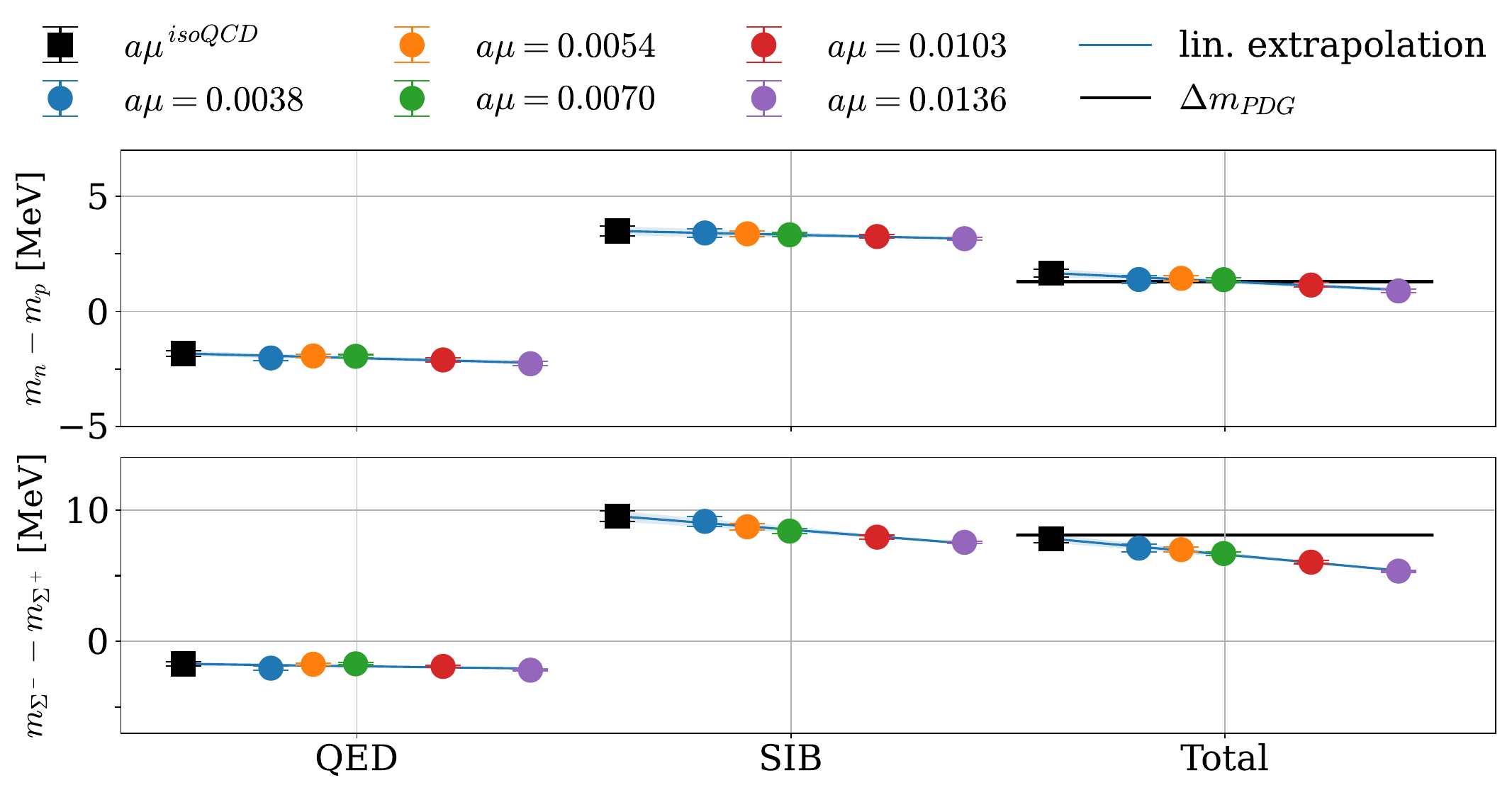}
\captionof{figure}{Extrapolation to the physical light quark mass for the nucleon and $\Sigma$ baryon, computed using the D96 ensemble. The individual QED and SIB contributions, together with the total mass splitting, are shown. Results at fixed valence-quark masses are displayed in different colours, together with the corresponding extrapolation bands (light blue) and the extrapolated values (black squares). The QED and SIB contributions are shown separately, as well as their sum ,denoted by total.
\label{fig:splitting}}
\end{center}

For each mass splitting, we extrapolate the results obtained at fixed valence-quark mass to the isoQCD light-quark mass $\mu^{\iso}$, treating the QED and SIB contributions separately. At the present stage of the analysis, we use a linear extrapolation based on all available valence-quark masses. In Fig.~\ref{fig:splitting}, we show the results for the nucleon and $\Sigma$ baryon using the B64 ensemble. 

\begin{figure}[t]
\centering
\includegraphics[width=\linewidth]{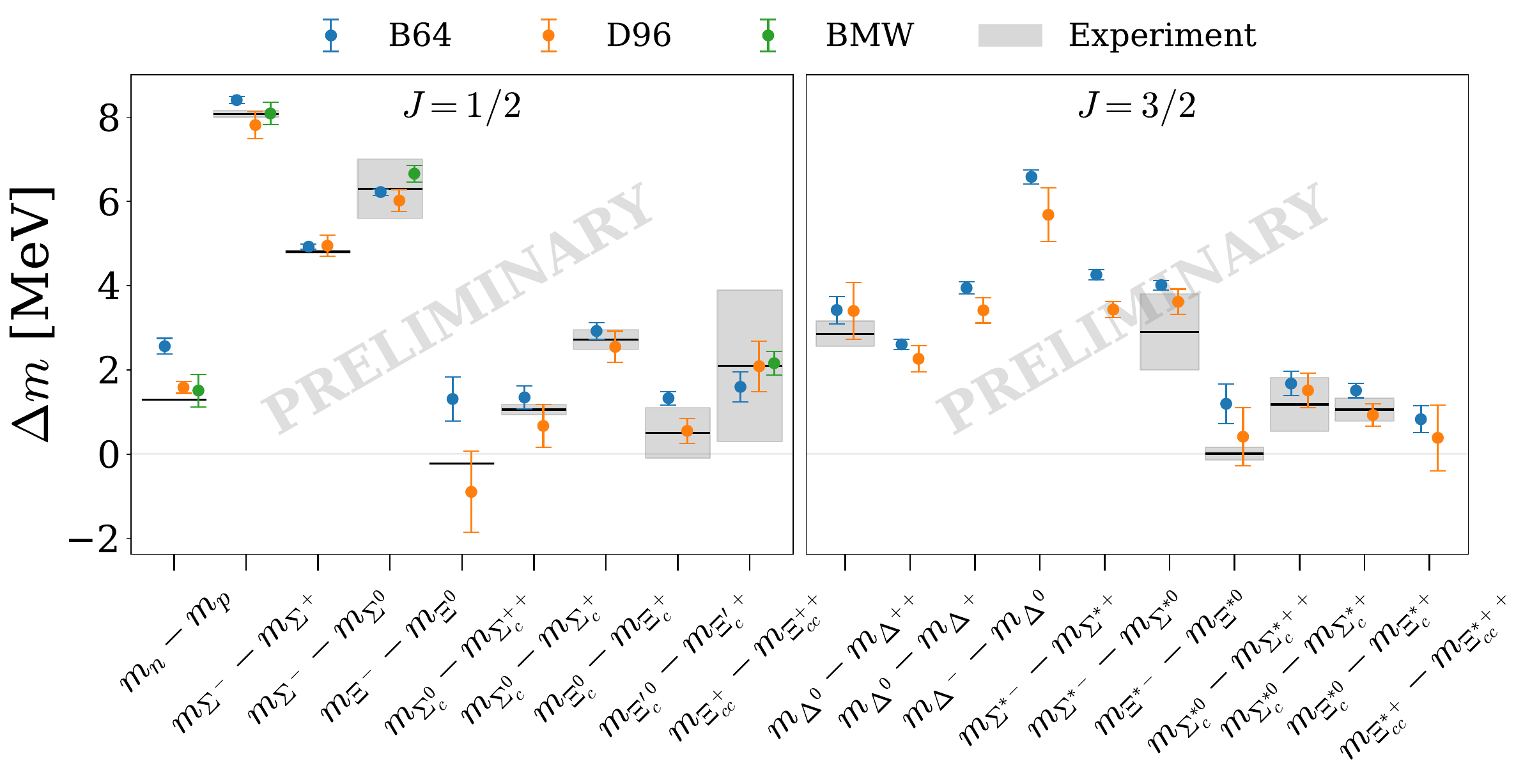}
\vspace{-0.8cm}
\captionof{figure}{Preliminary results for baryon mass splittings in the the  spin-$1\slash 2$ and spin-$3\slash 2$ .channels.}
\label{fig:PrelimResults}
\vspace{-0.3cm}
\end{figure}

We perform this analysis for the  mass splittings of SU(2) multiplets of spin-$1/2$ and spin-$3/2$ baryons composed of $u$, $d$, $s$, and $c$ quarks. The complete set of splittings considered is listed in Table~\ref{tab:splittings}, together with the corresponding experimental values and our preliminary results from the D96 ensemble. Results from both ensembles are also shown in Fig.~\ref{fig:PrelimResults}, where they are compared with their experimental values and with the results obtained by the BMW collaboration using $N_{\!f}{=}1{+}1{+}1{+}1$ QCD+QED ensembles~\cite{BMW:2014pzb}.
In general, we observe good qualitative agreement. A quantitative assessment, however, will be carried out after we finalise our  analysis, which will incorporate additional ensembles and a more comprehensive account of systematic uncertainties.
Such  a careful assessment of discretisation and other systematic effects is essential before a quantitative comparison with experimental values or for  reliable predictions. 
We observe a pronounced cancellation between the QED and SIB contributions in the charm sector, for both spin-$1/2$ and spin-$3/2$ baryons, leading to comparatively smaller total mass splittings than those observed for other baryons. In the latter case, the QED contribution appears to be subleading compared to the SIB contribution.\vspace*{-0.3cm}

\begin{table}[ht]
\centering
\small
\vspace{-0.3cm}
\renewcommand{\arraystretch}{1.4}
\begin{tikzpicture}
  \node[inner sep=0pt] (table) {
\begin{tabular}{|c|c|c|l|l|l|l|}
\cline{4-6}
\multicolumn{3}{c}{} & \multicolumn{3}{|c|}{\cellcolor{llbl} \textbf{Preliminary results on D96}} & \multicolumn{1}{c}{} \\
\cline{2-7}
\multicolumn{1}{c|}{}&\cellcolor{llbl} \textbf{Multiplet} &\cellcolor{llbl} \textbf{Splitting} &\cellcolor{llbl} \textbf{QED\,[MeV]} &\cellcolor{llbl} \textbf{SIB\,[MeV]} &\cellcolor{llbl}  \textbf{Sum\,[MeV]} &\cellcolor{llbl} \textbf{Exp.~[MeV]} \\
\hline
\cellcolor{llbl} & $N$ & $m_n - m_p$ & $-1.83(12)$ & $\pp 3.50(22)$ & $\pp 1.67(18)$ & $\pp 1.2933325(4)$ \\

\cline{2-7}
\cellcolor{llbl} & \multirow{2}{*}{$\Sigma$} 
& $m_{\Sigma^-} - m_{\Sigma^+}$ & $-1.71(16)$ & $\pp 9.53(42)$ & $\pp 7.81(32)$ & $\pp 8.08(8)$ \\\lcline{3-7}
\cellcolor{llbl} & & $m_{\Sigma^-} - m_{\Sigma^0}$ & $\pp 0.02(19)$ & $\pp 4.93(28)$ & $\pp 4.95(25)$ & $\pp 4.807(35)$ \\
\cline{2-7}
\cellcolor{llbl} & $\Xi$ & $m_{\Xi^-} - m_{\Xi^0}$ & $\pp 0.15(13)$ & $\pp 5.87(28)$ & $\pp 6.02(27)$ & $\pp 6.3(7)$ \\
\cline{2-7}
\cellcolor{llbl} & \multirow{2}{*}{$\Sigma_c$} 
& $m_{\Sigma_c^{0}} - m_{\Sigma_c^{++}}$ & $-8.16(74)$ & $\pp 7.27(63)$ & $-0.90(97)$ & $-0.220(13)$ \\\lcline{3-7}
\cellcolor{llbl} & & $m_{\Sigma_c^{0}} - m_{\Sigma_c^{+}}$ & $-2.81(37)$ & $\pp 3.48(40)$ & $\pp 0.67(51)$ & $\pp 1.06(12)$ \\
\cline{2-7}
\cellcolor{llbl} & $\Xi_c$ & $m_{\Xi_c^{0}} - m_{\Xi_c^{+}}$ & $-1.95(33)$ & $\pp 4.50(23)$ & $\pp 2.55(37)$ & $\pp 2.72(23)$ \\
\cline{2-7}
\cellcolor{llbl} & $\Xi'_c$ & $m_{\Xi_c^{\prime0}} - m_{\Xi_c^{\prime+}}$ & $-2.68(22)$ & $\pp 3.24(24)$ & $\pp 0.55(30)$ & $\pp 0.5(6)$ \\
\cline{2-7}
 \parbox[t]{2mm}{\multirow{-9}{*}{\cellcolor{llbl}\rotatebox[origin=c]{90}{\textbf{Spin-1/2 multiplets}}}} & $\Xi_{cc}$ & $m_{\Xi_{cc}^{+}} - m_{\Xi_{cc}^{++}}$ & $\pp 5.33(57)$ & $-3.24(24)$ & $\pp 2.09(60)$ & $\pp 2.1(1.8)$  \\
\hline
\hline
\cellcolor{llbl} & \multirow{3}{*}{$\Delta$} 
& $m_{\Delta^{0}} - m_{\Delta^{++}}$ & $-4.57(61)$ & $\pp 7.97(47)$ & $\pp 3.40(68)$ & $\pp 2.86(30)$\\\lcline{3-7}
\cellcolor{llbl} & & $m_{\Delta^{0}} - m_{\Delta^{+}}$  & $-1.69(12)$ & $\pp 3.95(34)$ & $\pp 2.26(32)$ & \pp --- \\\lcline{3-7}
\cellcolor{llbl} & & $m_{\Delta^{-}} - m_{\Delta^{0}}$  & $-0.73(15)$ & $\pp 4.14(30)$ & $\pp 3.42(31)$ & \pp --- \\

\cline{2-7}
\cellcolor{llbl} & \multirow{2}{*}{$\Sigma^*$} 
& $m_{\Sigma^{*-}} - m_{\Sigma^{*+}}$ & $-1.75(20)$ & $\pp 7.44(74)$ & $\pp 5.68(64)$ & \pp ---\\\lcline{3-7}
\cellcolor{llbl} & & $m_{\Sigma^{*-}} - m_{\Sigma^{*0}}$ & $-0.34(11)$ & $\pp 3.78(22)$ & $\pp 3.43(19)$ & \pp ---\\

\cline{2-7}
\cellcolor{llbl} & $\Xi^*$ & $m_{\Xi^{*-}} - m_{\Xi^{*0}}$ & $-0.04(21)$ & $\pp 3.66(35)$ & $\pp 3.62(31)$ & $\pp 2.9(9)$ \\
\cline{2-7}
\cellcolor{llbl} & \multirow{2}{*}{$\Sigma^*_c$} 
& $m_{\Sigma^{*0}_c} - m_{\Sigma^{*++}_c}$ & $-6.19(54)$ & $\pp 6.61(48)$ & $\pp 0.41(69)$ & $\pp 0.01(15)$ \\\lcline{3-7}
\cellcolor{llbl} & & $m_{\Sigma^{*0}_c} - m_{\Sigma^{*+}_c}$ & $-2.01(31)$ & $\pp 3.53(35)$ & $\pp 1.51(42)$ & $\pp 1.18(64)$ \\

\cline{2-7}
\cellcolor{llbl} & $\Xi^*_c$ & $m_{\Xi^{*0}_c} - m_{\Xi^{*+}_c}$ & $-2.36(17)$ & $\pp 3.29(22)$ & $\pp 0.93(27)$ & $\pp 1.06(27)$  \\
\cline{2-7}
\parbox[t]{2mm}{\multirow{-10}{*}{\cellcolor{llbl}\rotatebox[origin=c]{90}{\textbf{Spin-3/2 multiplets}}}}& $\Xi^*_{cc}$ & $m_{\Xi_{cc}^{*+}} - m_{\Xi_{cc}^{*++}}$ & $\pp 3.84(62)$ & $-3.45(49)$ & $\pp 0.39(79)$ & \pp --- \\

\hline
\end{tabular}%
  };
  \begin{scope}[on background layer]
  \node[
    rotate=38,
    scale=6.5,
    text opacity=0.2,
    text=gray,
    font=\bfseries
  ] at ([xshift=4mm,yshift=-6mm]table.center) {PRELIMINARY};
  \end{scope}
\end{tikzpicture}
\vspace{-1.4cm}
\captionof{table}{Preliminary results for the isospin mass splittings of the analysed SU(2) spin-$1/2$ and spin-$3/2$ multiplets, computed on the D96 ensemble. Experimental values are taken from the PDG~\cite{ParticleDataGroup:2026mpi}, except for the recent LHCb measurement of the $\Xi_{cc}^{+}$ mass splitting~\cite{LHCb:2026pxn}.}\vspace*{-0.3cm}
\label{tab:splittings}
\end{table}

\section{Conclusions and outlook}
A first analysis of the octet and decuplet baryon mass splittings arising from SU(2) isospin breaking is presented. Using two $N_{\!f}{=}2{+}1{+}1$ twisted mass ensembles we demonstrate the applicability of the RM123 method in the determination of mass splitings. Currently only statistical uncertainties are evaluated and no continuum limit has been carried out. Nevertheless, there is reasonable agreement with experimentally determined mass splittings. 

We plan to analyse more ensembles at additional lattice spacings to enable us to take the continuum limit, after  evaluating  systematic uncertainties on e.g. the fitting procedure.  Beyond  correcting for the universal finite-volume contributions, we plan  to assess finite-volume effects using ensembles with larger spatial volumes. 
An important point in this regard is that, while for spin-1/2 baryons we can smoothly take the infinite-volume limit, the situation is more subtle for spin-3/2 baryons. As the volume is increased, the relevant energy levels may move above threshold and receive contributions from the corresponding low-lying meson-baryon states. We will assess the consequence of this and examine how to  reliably isolate the baryon states of interest via e.g. performing a a generalized eigenvalue problem analysis.
A  complete determination of the leading-order corrections,  should also include an assessment of electromagnetic unquenching effects by including electromagnetic interactions between the sea and valence quarks. This will be undertaken in future work.

\setlength{\parskip}{0pt}
\section*{Acknowledgements}
{
We thank all members of the ETM Collaboration for the most enjoyable collaboration. C.A., S.B. A.E. and C.I. acknowledge support from the projects EXCELLENCE/0524/0459 (IMAGE-N), EXCELLENCE/0524/0017 (MuonHVP), POST-DOC/0925-ROandOther/0148  (HiQO), EXCELLENCE/0524/0455 (DeNuTra), respectively, implemented under the programme of social cohesion “THALIA 2021-2027” co-funded by the European Union and the Republic of Cyprus, through the Research and Innovation Foundation. S.B. also acknowledges support by the project OPPTY-ERC/0524/0006 (QWERTY), funded by the Republic of Cyprus, through the Research and Innovation Foundation. C.S. is supported under the AQTIVATE EJD from the European Union’s research and innovation programme under the Marie Sklodowska-Curie Doctoral Networks action and Grant Agreement No 101072344.  We  gratefully acknowledge the EuroHPC Joint Undertaking for granting the project ID EHPC-EXT-2025E02-110 access to Leonardo at CINECA and the project ID  EHPC-REG-2026R01-195 access to Jupyter at JSC.
}

\bibliography{references.bib}

\providecommand{\href}[2]{#2}\begingroup\raggedright\begin{thebibliography}{10}

\bibitem{deDivitiis:2011eh}
G.~M. de~Divitiis et~al., \emph{{Isospin breaking effects due to the up-down mass difference in Lattice QCD}}, \href{http://dx.doi.org/10.1007/JHEP04(2012)124}{\emph{JHEP} {\bf 04} (2012) 124}, [\href{http://arxiv.org/abs/1110.6294}{{\tt 1110.6294}}].

\bibitem{deDivitiis:2013xla}
{\scshape RM123} collaboration, G.~M. de~Divitiis, R.~Frezzotti, V.~Lubicz, G.~Martinelli, R.~Petronzio, G.~C. Rossi et~al., \emph{{Leading isospin breaking effects on the lattice}}, \href{http://dx.doi.org/10.1103/PhysRevD.87.114505}{\emph{Phys. Rev. D} {\bf 87} (2013) 114505}, [\href{http://arxiv.org/abs/1303.4896}{{\tt 1303.4896}}].

\bibitem{Giusti:2017dmp}
D.~Giusti, V.~Lubicz, C.~Tarantino, G.~Martinelli, F.~Sanfilippo, S.~Simula et~al., \emph{{Leading isospin-breaking corrections to pion, kaon and charmed-meson masses with Twisted-Mass fermions}}, \href{http://dx.doi.org/10.1103/PhysRevD.95.114504}{\emph{Phys. Rev. D} {\bf 95} (2017) 114504}, [\href{http://arxiv.org/abs/1704.06561}{{\tt 1704.06561}}].

\bibitem{FlavourLatticeAveragingGroupFLAG:2024oxs}
{\scshape Flavour Lattice Averaging Group (FLAG)} collaboration, Y.~Aoki et~al., \emph{{FLAG review 2024}}, \href{http://dx.doi.org/10.1103/nfzp-p5dn}{\emph{Phys. Rev. D} {\bf 113} (2026) 014508}, [\href{http://arxiv.org/abs/2411.04268}{{\tt 2411.04268}}].

\bibitem{ExtendedTwistedMass:2024nyi}
{\scshape Extended Twisted Mass} collaboration, C.~Alexandrou et~al., \emph{{Strange and charm quark contributions to the muon anomalous magnetic moment in lattice QCD with twisted-mass fermions}}, \href{http://dx.doi.org/10.1103/PhysRevD.111.054502}{\emph{Phys. Rev. D} {\bf 111} (2025) 054502}, [\href{http://arxiv.org/abs/2411.08852}{{\tt 2411.08852}}].

\bibitem{Hayakawa:2008an}
M.~Hayakawa and S.~Uno, \emph{{QED in finite volume and finite size scaling effect on electromagnetic properties of hadrons}}, \href{http://dx.doi.org/10.1143/PTP.120.413}{\emph{Prog. Theor. Phys.} {\bf 120} (2008) 413--441}, [\href{http://arxiv.org/abs/0804.2044}{{\tt 0804.2044}}].

\bibitem{Davoudi:2018qpl}
Z.~Davoudi, J.~Harrison, A.~J{\"u}ttner, A.~Portelli and M.~J. Savage, \emph{{Theoretical aspects of quantum electrodynamics in a finite volume with periodic boundary conditions}}, \href{http://dx.doi.org/10.1103/PhysRevD.99.034510}{\emph{Phys. Rev. D} {\bf 99} (2019) 034510}, [\href{http://arxiv.org/abs/1810.05923}{{\tt 1810.05923}}].

\bibitem{Evangelista:2025dzi}
{\scshape Extended Twisted Mass} collaboration, A.~Evangelista, S.~Bacchio, R.~Frezzotti, G.~Gagliardi, M.~Garofalo, N.~Kalntis et~al., \emph{{Valence leading isospin breaking contributions to $a_{\mu}^{\mathrm{HVP-LO}}$}}, \href{http://dx.doi.org/10.22323/1.466.0260}{\emph{PoS} {\bf LATTICE2024} (2025) 260}, [\href{http://arxiv.org/abs/2501.19350}{{\tt 2501.19350}}].

\bibitem{Frezzotti:2000nk}
{\scshape Alpha} collaboration, R.~Frezzotti, P.~A. Grassi, S.~Sint and P.~Weisz, \emph{{Lattice QCD with a chirally twisted mass term}}, \href{http://dx.doi.org/10.1088/1126-6708/2001/08/058}{\emph{JHEP} {\bf 08} (2001) 058}, [\href{http://arxiv.org/abs/hep-lat/0101001}{{\tt hep-lat/0101001}}].

\bibitem{Frezzotti:2003ni}
R.~Frezzotti and G.~C. Rossi, \emph{{Chirally improving Wilson fermions. 1. O(a) improvement}}, \href{http://dx.doi.org/10.1088/1126-6708/2004/08/007}{\emph{JHEP} {\bf 08} (2004) 007}, [\href{http://arxiv.org/abs/hep-lat/0306014}{{\tt hep-lat/0306014}}].

\bibitem{Frezzotti:2016lwv}
R.~Frezzotti, G.~Rossi and N.~Tantalo, \emph{{Sea quark QED effects and twisted mass fermions}}, \href{http://dx.doi.org/10.22323/1.256.0320}{\emph{PoS} {\bf LATTICE2016} (2016) 320}, [\href{http://arxiv.org/abs/1612.02265}{{\tt 1612.02265}}].

\bibitem{BMW:2014pzb}
{\scshape BMW} collaboration, S.~Borsanyi et~al., \emph{{Ab initio calculation of the neutron-proton mass difference}}, \href{http://dx.doi.org/10.1126/science.1257050}{\emph{Science} {\bf 347} (2015) 1452--1455}, [\href{http://arxiv.org/abs/1406.4088}{{\tt 1406.4088}}].

\bibitem{ParticleDataGroup:2026mpi}
{\scshape Particle Data Group} collaboration, F.~Takahashi et~al., \emph{{Review of Particle Physics}}, \href{http://dx.doi.org/10.1142/s0217751x26300115}{\emph{Int. J. Mod. Phys. A} {\bf 41} (2026) 2630011}.

\bibitem{LHCb:2026pxn}
{\scshape LHCb} collaboration, R.~Aaij et~al., \emph{{Observation of the Doubly Charmed Baryon {\ensuremath{\Xi}}cc+ with the LHCb Run 3 Detector}}, \href{http://dx.doi.org/10.1103/dmv6-7gdv}{\emph{Phys. Rev. Lett.} {\bf 137} (2026) 021902}, [\href{http://arxiv.org/abs/2603.28456}{{\tt 2603.28456}}].

\end{thebibliography}\endgroup
\bibliographystyle{JHEP}

\end{document}